\documentclass[a4,12pt]{article}

\usepackage{hyperref}
\usepackage{amsmath}
\usepackage{graphicx}
\usepackage{color}
\usepackage{amssymb} 
\newcommand{\Mat}[1]{{{\boldsymbol{#1}}}}
\newcommand{\abs}[1]{\left\vert#1\right\vert}
\def\be{\begin{equation}}
\def\ee{\end{equation}}
\def\bea{\begin{eqnarray}}
\def\eea{\end{eqnarray}}
\def\dd{\mathrm{d}}
\title{\bf Equations of motion of the mass centers in a scalar theory of gravity with a preferred frame}
\author{
Mayeul Arminjon\\
\small\it Univ. Grenoble Alpes, CNRS, Grenoble INP
, 3SR, F-38000 Grenoble, France\\
\small\it  E-mail: Mayeul.Arminjon@3sr-grenoble.fr.
}
\date{}

\begin{document}

\maketitle
\begin{abstract}
The theory considered interprets gravity as a pressure force. Thus, the scalar gravitational field defines the gravity acceleration field. However, it also determines the relation between the flat ``background metric'' and a curved ``physical metric''. Here we derive the equations of motion of the mass centers of a system of weakly gravitating bodies in the second version of that theory. We use the framework which was built and used for the first version. Namely, we use an asymptotic scheme of post-Newtonian (PN) approximation to derive the local (field) PN equations, and by integration inside the bodies we deduce from those local equations the equations of motion of the mass centers, using also an asymptotic framework for the good separation between the different bodies.
\end{abstract}

{\renewcommand{\baselinestretch}{0.6}\small\tableofcontents}\normalsize

\section{Introduction}

The investigated theory starts from an interpretation of gravity as a pressure force in a perfect fluid or ``micro-ether'', of which the matter particles would be self-organized local flows, e.g. vortices: see Ref. \cite{A65} and references therein. Special relativity, in the version of Lorentz and Poincaré, keeps Lorentz's concept of the ether being a privileged inertial frame. This is compatible with the idea expressed in the first sentence: Lorentz's ether is defined by the averaged motion of the micro-ether. Metric modifications: gravitational space contraction and time dilation, are naturally associated with the heterogeneous pressure field that causes the gravitational force. This leads to a Riemannian space metric and to clock periods that depend on the time and the space position, that together make a Lorentzian spacetime metric \cite{A65}, as in general relativity (GR) and in many extensions of it. However, motion is defined by an extension to such a ``curved spacetime'' of Planck's relativistic modification of Newton's second law. 
The theory has been developed further, see Ref. \cite{A65} for a recent review including a rather detailed motivation. These developments include the dynamics of a continuous medium; coupling with Maxwell's electromagnetic field; coupling with quantum mechanics, in particular with the Dirac equation; basic cosmology in the framework of the homogeneous model of the Universe; and of course celestial mechanics.\\

For celestial mechanics, an asymptotic scheme of post-Newtonian (PN) approximation has been proposed within that theory \cite{A23}. That scheme differs from the standard scheme of PN approximation used (see e.g. Refs. \cite{Fock1964, Chandra1965, Weinberg1972, Kopeikin-Vlasov2004}) for GR --- e.g. in that all fields, including the matter fields, are expanded as function of the small parameter, as is necessary to obtain asymptotic expansions, and not merely the gravitational field as is done in the standard scheme.
\footnote{\
See Ref. \cite{A36} for a detailed comparison of the two schemes within GR, including a comparison of the global equations of motion derived with either the asymptotic scheme or the standard one. That reference contains also a discussion of more recent works on the PN approximation of GR.
}
The asymptotic PN scheme has allowed us to derive the equations of motion of the mass centers of a weakly-gravitating system of bodies like our solar system within the investigated theory \cite{A25-26}. These equations of motion have been slightly corrected thanks to the introduction of an asymptotic formalism for the ``good separation'' of the celestial bodies \cite{A32}. In a later step, it has been found that the initial version (``v1'') of the theory led to a violation of the weak equivalence principle at the point particle limit, that was due to the anisotropic character of the spatial metric (a character that is shared with the standard form of the Schwarzschild metric of general relativity). A new version (``v2'') of the theory has been built, in which the spatial metric is (pointwise) isotropic and, therefore, that violation does not exist any more in that new version \cite{A35}. \\

The asymptotic post-Newtonian scheme has been adapted to get the local PN equations of motion in v2 \cite{A35}. Also, the general form of the equations of motion of the mass centers in v2 has been obtained \cite{A35}, Eqs. (\ref{delta-a}) and (\ref{masscent-ord1}) here. However, the explicit form of the latter equations has not been derived yet. It is the purpose of the present paper to derive this fully-detailed form, in order to be able later to implement them in a software for numerical calculations of celestial mechanics. (As with other relativistic theories of gravitation, and in contrast with Newtonian celestial mechanics where analytical solutions can be found in important situations (see e.g. \cite{González2022a, González2022b, González2023, Zhivkov-Mladenov2024} for recent findings), the greater complexity of this theory as compared with Newtonian theory enforces one to have recourse to numerical calculations to solve the equations of motion.) 
The next three sections provide a summary of previous work, that makes the paper self-contained. After them, Sect. \ref{EMMC} contains the main results of the present work, referring to Appendix \ref{Integrals-eta} for detailed auxiliary calculations. 
Section \ref{Conclusion} presents our conclusions.

\section{Main Equations of the Scalar Theory (V2)}

See Ref. \cite{A65} for a more detailed summary of the theory, including references. The preferred reference frame $\mathcal{E}$ 
assumed by the theory is an inertial frame for the flat metric $\Mat{\gamma }^0$. 
It means that there are spacetime coordinates $x^\mu \ (\mu =0,...,3)$ which are both {\it adapted} to $\mathcal{E}$ (i.e., any point bound to the frame $\mathcal{E}$ has constant spatial coordinates $x^i\ (i=1,2,3)$) and {\it Cartesian} for $\Mat{\gamma }^0$ --- i.e., $\gamma ^0_{\mu \nu }=\eta _{\mu \nu }$, where $\eta _{\mu \nu }\ (\mu ,\nu =0,...,3)$ are the standard components of the Minkowski metric, which make the matrix $\eta =(\eta _{\mu \nu }):=\mathrm{diag}(1,-1,-1,-1)$. Such coordinates allow one to define the inertial time in the frame $\mathcal{E}$, $T:=x^0/c$. Also, in such coordinates, the line elements of $\Mat{\gamma }^0$ and the ``physical" metric $\Mat{\gamma }$ are given respectively by
\bea \label{gamma^0}
(\dd s^0)^2 & := & \gamma^0_{\mu\nu} \dd x^\mu \dd x^\nu = (\dd x^0)^2 - \dd x^i \dd x^i \\ \label{gamma}
\dd s^2 & := & \gamma_{\mu\nu} \dd x^\mu \dd x^\nu = \beta^2 (\dd x^0)^2 - \beta ^{-2} \dd x^i \dd x^i
\eea
with $\beta$ the scalar gravitational field. Thus, the spatial metric $\Mat{g}$ in the frame $\mathcal{E}$ (the spatial part \cite{L&L, Moller1952, Cattaneo1958}, in the frame $\mathcal{E}$, of the ``physical'', curved metric $\Mat{\gamma }$) has components
\be
g_{ij}=\beta^{-2} \,g^0_{\,ij}
\ee
where the $g^0_{ij}$'s are the components of the Euclidean spatial metric $\Mat{g}^0$  (the spatial part, in the frame $\mathcal{E}$, of the flat metric $\Mat{\gamma }^0$), with $g^0_{\,ij}=\delta_{\,ij}$ in Cartesian coordinates. The field $\beta$ also generates a gravity acceleration, which enters Newton's second law for a point particle. The dynamical equation for a continuous medium, inferred from that, is
\begin{equation} \label{continuum}
T_{\mu;\nu}^{\nu} = b_{\mu}.				
\end{equation}
Here ${\bf T}$ is the energy(-momentum-stress) tensor of matter and nongravitational fields, and $b_\mu$ is defined by
\begin{equation} \label{definition_b}
b_0(\mathbf{T}) := \frac{1}{2}\,g_{jk,0}\,T^{jk}, \quad b_i(\mathbf{T}) := -\frac{1}{2}\,g_{ik,0}\,T^{0k}.
\end{equation}
(Indices are raised and lowered with metric $\Mat{\gamma}$, unless mentioned otherwise. Semicolon means covariant differentiation using the Christoffel connection associated with metric $\Mat{\gamma}$.) Note that (\ref{continuum}) differs from the corresponding equation in GR and other ``metric theories" of gravitation, in which the r.h.s. is zero instead of $b_\mu$. (Well-known and excellent references for GR are the books \cite{Moller1952, L&L, Fock1964, Weinberg1972, Stephani}.) Finally, the equation for the scalar field $\psi :=-\mathrm{Log}\,\beta $ is \cite{A35}:
\be\label{field}
\square \psi  := \psi _{,0,0}-\Delta\, \psi  = \frac{4 \pi G}{c^2} \sigma.
\ee
Here $\Delta := \Delta_{\Mat{g}^0}$ is the usual Laplace operator, defined with the Euclidean space metric $\Mat{g}^0$ (thus $\Delta_{\Mat{g}^0}\, \psi = \psi_{,i,i}$ in Cartesian coordinates for $\Mat{g}^0$, i.e. such that $g^0_{i j}=\delta _{i j}$). $G$ is Newton's gravitation constant. And
\be\label{sigma}
\sigma :=T^{0 0}
\ee
is the energy component of the total energy tensor ${\bf T}$ of matter and non-gravita-\\
tional fields, in coordinates adapted to the reference frame $\mathcal{E}$, and with the time coordinate being $x^0=cT$, with $T$ the inertial time in $\mathcal{E}$. We take ${\bf T}$ in mass units, hence it is $c^2 \sigma $ which is really an energy density. It follows from (\ref{field}) that
\be \label{def_V}
V := c^2 \psi
\ee
satisfies the wave equation with the same r.h.s. as Poisson's equation for the Newtonian potential $U_\mathrm{N}$. Since the retardation effects should become negligible in the Newtonian limit, $V$ is a natural equivalent of $U_\mathrm{N}$.

\section{Asymptotic Scheme of PN Approximation and Local PN Equations}\label{PNA}

In order to obtain asymptotic expansions of the unknown fields: pressure, velocity, scalar gravitational field, etc., we conceptually consider a family of gravitational systems, $(\mathrm{S}_\lambda)$, indexed by the field-strength parameter $\lambda := \mathrm{Sup}\, \psi $.
\footnote{\
According to Eq. (\ref{gamma}), a weak gravitational field may be characterized in this theory by $\beta\simeq 1$, thus $\psi=-\mathrm{Log}\,\beta \simeq 0$. Moreover, since $\beta = \exp (-\psi)$, we have $0\le \beta$, and from the retarded solution of (\ref{field}) with $\sigma \ge 0$ we have $\psi \ge 0$, hence $\beta \le 1$. Therefore, the maximum strength of the gravitational field is characterized by $\mathrm{Inf}\, \beta $, or equivalently by $\mathrm{Sup}\, \psi $.
}
The real system of interest, $\mathrm{S}$ (e.g. the solar system), is assumed to belong to the family for a small value $\lambda_0 \ll 1$ of that parameter, thus $\mathrm{S}= \mathrm{S}_{\lambda _0}$, so that we have indeed a weakly-gravitating system. The family $(\mathrm{S}_\lambda)$ is defined by a family of specific initial conditions \cite{A23,A35}. We adopt $\lambda $-dependent units of time and mass: $[\mathrm{M}]_\lambda = \lambda[\mathrm{M}]$ and $[\mathrm{T}]_\lambda = [\mathrm{T}]/\sqrt{\lambda}$ for system $\mathrm{S}_\lambda$, where $[\mathrm{M}]$ and $[\mathrm{T}]$ are the starting units of mass and time. Then all fields become order $\lambda^0$, and the small parameter $\lambda$ is proportional to $1/c^2$ (indeed $\lambda=(c_0/c)^2$, where $c_0$ is the velocity of light in the starting units). This makes it easy to derive asymptotic expansions. \\

In this work also, we will consider a barotropic perfect fluid. Thus, the proper rest-mass density $\rho^\ast$ depends only on the pressure $p$:
\be\label{barotropic}
\rho^\ast= F(p).
\ee
Moreover, we will still consider only the {\it first post-Newtonian (1PN) approximation,} that immediately follows the Newtonian approximation. One begins with expansions for the independent fields: $V$, the pressure $p$, and the velocity ${\bf u}:=\dd {\bf x}/\dd T$:
\be \label{expans_base}
V= V_0 + V_1/c^2 + O(c^{-4}),\quad p= p_0 + p_1/c^2 + O(c^{-4}), \quad {\bf u}= {\bf u}_0 + {\bf u}_1/c^2 + O(c^{-4})
\ee
and one deduces expansions for the other fields \cite{A35}. Entering these expansions into the field equations (\ref{continuum}) and (\ref{field}), using the standard expression of ${\bf T}$ for a perfect fluid, and identifying the powers of $1/c^2$, one gets the expansions of the field equations \cite{A35}. In particular, the equation (\ref{field}) for the scalar field expands to
\be \label{expans_fieldeq}
\Delta V_0 = -4\pi G \sigma _0, \qquad \Delta V_1 = -4\pi G \sigma _1 + \partial_T^2 V_0
\ee
where $\sigma = \sigma _0 + \sigma _1/c^2 +O(c^{-4})$ is the 1PN expansion of the active mass density (\ref{sigma}). It follows \cite{A35} that $U := V_0$ is the Newtonian potential associated with $\sigma _0$:
\be \label{U}
U := V_0 = \mathrm{N.P.}[\sigma _0] \qquad \left(\mathrm{N.P.}[\tau](\mathbf{X},T) := G \int  \tau (\mathbf{x},T)\dd {\mathsf V}(\mathbf{x})/\abs{\mathbf{X - x}}\right)
\ee
and that
\be \label{V_1}
V_1 = B + \frac{\partial ^2 W}{\partial T^2}, \qquad B := \mathrm{N.P.}[\sigma_1]
\ee
with
\be\label{W}
  W(\mathbf{X},T) := G \int \abs{\mathbf{X - x}}
 \sigma _0(\mathbf{x},T)\dd {\mathsf V}(\mathbf{x})/2.
\ee
Other important expansions are those for $\rho $, the density of rest-mass in the preferred frame and with respect to the Euclidean volume measure $\dd {\mathsf V}:= \sqrt{g^0}\, \dd^3 {\bf x}$ (with $g^0=\mathrm{det}\,(g^0_{ij})$, $g^0=1$ in Cartesian coordinates), and for the proper rest-mass density $\rho ^\ast$. The density $\rho $ is related to $\rho ^\ast$ by Lorentz and gravitational contraction. We have \cite{A23,A35}
\be \label{expans_matter_0}
\rho ^\ast _0= F(p_0)=\rho _0 = \sigma _0
\ee
and
\be \label{rho_1}
 \rho ^\ast_1 = F'(p_0).p_1,\quad \rho_1 = \rho ^\ast_1 + \rho_0 \left(\frac{{\bf u}_0^2}{2} + 3U \right)
\ee
\be \label{sigma_1}
\sigma_1 = \rho_1 + \rho_0 \left(\frac{{\bf u}_0^2}{2} -U + \Pi_0 \right).
\ee
(Here $\Pi_0$ is the first, zero-order term, in the expansion of the density of elastic energy per unit rest mass $\Pi$.) The equation of motion for these PN fields is, at the order zero:
\begin{equation}\label{i-ord0}
    \partial _T (\rho_0 u_0^i) + \partial _j (\rho_0 u_0^i u_0^j)=\rho_0
    U_{,i}-p_{0,i}
\end{equation}
which is just the Newtonian equation of motion for a perfect fluid. It is completed by the continuity equation that expresses mass conservation at the order zero:
\begin{equation}\label{T-ord0}
  \partial _T \rho_0 + \partial _j (\rho_0 u_0^j) = 0.
\end{equation}
(With this, equation (\ref{i-ord0}) implies Euler's equation.) The local equation of motion of the order one in $1/c^2$ is:
\begin{eqnarray}\label{i-ord1}
    \partial _T (\rho_0 u_1^i + \theta_1 u_0^i) + \partial _j ( \rho_0 u_0^i u_1^j + \rho_0 u_1^i u_0^j +\theta_1  u_0^i u_0^j ) - \rho_0 (u_0^i \partial_T U + {\bf u}_0^2 \partial_i U) & & \nonumber \\ = \sigma_1 U_{,i} + \rho_0 V_{1,i}+ p_0 U_{,i} -2U p_{0,i}-p_{1,i} &  &
\end{eqnarray}
with \cite{A35}
\be\label{theta1}
\theta_1 = \sigma_1 + p_0 + 4 \rho_0 U =\rho_1 + \rho_0 \left(\frac{{\bf u}_0^2}{2} +3U + \Pi_0 \right) +p_0.
\ee
It is completed by the following equation:
\begin{equation}\label{mass-ord1}
  \partial _T \rho_1 + \partial _j(\rho_1 u_0^j + \rho_0 u_1^j)=0
\end{equation}
that, together with (\ref{T-ord0}), says that mass is conserved at the first PNA of that theory.

\section{Asymptotic Framework for the Good Separation}

We conceptually consider a family $(\mathrm{S}'^\eta)$ of PN gravitating systems, i.e., of systems governed by the 1PN equations listed in Sect. \ref{PNA}, with $\mathrm{S}'^{\eta_0}$ being the 1PN description $\mathrm{S}'$ of the real system $\mathrm{S}$ \cite{A32}. Let the system S be made of $N$ massive bodies; let ${\bf a}_0$ be the zero-order position in the frame $\mathcal{E}$ of the center of mass of body $(a)\ (a=1,...,N)$, and let $r_a$ be its radius. To account for the good separation of the $N$ bodies, we introduce the separation parameter \cite{A25-26}
\begin{equation}\label{eta}
  \eta_0 := \max_{a \neq b}(r_b/\abs{{\bf a}_0-{\bf b}_0})
\end{equation}
and we assume $\eta_0 \ll 1$. The family $(\mathrm{S}'^\eta)$ is defined by a family of initial conditions \cite{A32}, ensuring that
\begin{equation}\label{a-eta-b-eta}
  (r_{ab})^\eta := \abs{\mathbf{a}_0^{\eta}-\mathbf{b}_0^{\eta}}=
  \mathrm{ord}(\eta^{-1}) \quad \mathrm{for\ }a \neq b.
\end{equation}
We assume that the bodies themselves do not depend on the separation parameter $\eta$ (for this we define the initial zero-order density field inside each body to be independent of $\eta$ \cite{A32}). And we assume \cite{A32} that, consistently with (\ref{a-eta-b-eta}) and the Newtonian estimate
\be \label{adot-estimate}
\dot{{\bf a}}_0^2 \approx 2 GM^0_N/r^0_{aN}
\ee
(valid when body $(N)$ is dominating), the zero-order translation velocities of the mass centers in system $\mathrm{S}'^\eta$ vary like $\eta^{1/2}$:
\begin{equation}\label{adot-eta}
  (\dot{a}_0^i)^\eta = \mathrm{ord}(\eta^{1/2}).
\end{equation}
(Here and in the sequel, the upper dot means $\dd/\dd T$.) We also assume that the bodies undergo rigid rotation at the Newtonian approximation, which is correct in order to compute the equations of motion up to the order $\eta^3$ included \cite{A36}: when the point ${\bf x}$ belongs to body $(a)$
\be\label{u-rigid}
  u_0^i =  \dot{a}_0^i + \Omega^{(a)}_{ji}(x^j - a_0^j ),\ (\Omega^{(a)}_{ji} +\Omega^{(a)}_{ij} = 0),\quad \mathrm{or}\quad
  \mathbf{u}_0 =\dot{\mathbf{a}}_0+ \boldsymbol{\omega}_{a}\wedge(\mathbf{x}-{\bf a}_0).
\ee
Moreover, the rotation velocities are assumed \cite{A32} to verify 
\begin{equation}\label{omega-eta}
  (\Omega^{(a)}_{ji})^\eta = o(\eta^{1/2}).
\end{equation}

\section{Equations of Motion of the Mass Centers}\label{EMMC}

We define the exact masses and mass centers through the rest-mass density $\rho$
\begin{equation}\label{defmasscent}
  M_a :=\int_{\mathrm{D}_a}\rho\, \dd V,\qquad \mathbf{a} := \frac{1}{M_a}\int_{\mathrm{D}_a}\rho\mathbf{x}\,\dd V
\end{equation}
where $\mathrm{D}_a$ is the (time-dependent) domain made of the spatial positions ${\bf x} := (x^i)$ of the particles constituting
body $(a)$ ($a=1,..., N$) in the reference frame $\mathcal{E}$. At the 1PN approximation, the mass and the mass center are approximated by \cite{A25-26}
\begin{equation}\label{defPNmass}
  M_a^{(1)}=M^0_a+M_a^1/c^2,\qquad M^0_a:= \int_{\mathrm{D}_a}\rho_0
  \,\dd V,\qquad M_a^1:=\int_{\mathrm{D}_a}\rho_1 \,\dd V
\end{equation}
\begin{equation}\label{defPNmasscent}
  \mathbf{a}_{(1)}:= \frac{1}{M_a^{(1)}}\int_{\mathrm{D}_a}\rho_{(1)}\mathbf{x}\,\dd V=
  \frac{1}{M_a^{(1)}} \left (M^0_a\,\mathbf{a} _{0} +M_a^{1}\,\mathbf{a} _{1} /c^2 \right)
\end{equation}
with
\begin{equation}\label{defmasscent-ord0-ord1}
  \mathbf{a}_{0} \, := \frac{1}{M^0_a} \int_{\mathrm{D}_a}\rho_0\mathbf{x}\,\dd V,\qquad \mathbf{a}_{1}:= \frac{1}{M_a^{1}}\int_{\mathrm{D}_a}\rho_{1}\mathbf{x}\,\dd V.
\end{equation}
Note that $M^0_a$ and $\mathbf{a}_{0} $ are the Newtonian mass and mass
center. It results from (\ref{T-ord0}) and (\ref{mass-ord1}) that $M^0_a$ and $M^1_a$ depend negligibly on time \cite{A25-26}. \\

{\it Henceforth, the index 0 will be omitted for the zero-order (Newtonian) quantities, for conciseness. } (Therefore, the exact quantities, if needed, should now be denoted by adding the subscript $_\mathrm{exact}$ to the initial notation, e.g. on the left-hand sides of Eq. (\ref{expans_base}), or in Eq. (\ref{defmasscent}).)
One finds from (\ref{defPNmass})-(\ref{defmasscent-ord0-ord1}) that the PN correction to the position of the mass center is given by \cite{A25-26}
\be\label{delta-a}
{\bf a}_{(1)} - {\bf a} = \frac{M_a^1}{c^2M_a}({\bf a}_1 - {\bf a}) + O(c^{-4}).
\ee
In the final equations of motion, we shall also use the notations
\begin{equation}\label{def-posi}
  \mathbf{x}_{a} := \mathbf{a}
\quad \mathbf{x}_{1a} :=
  c^2(\mathbf{a}_{(1)}-\mathbf{a})
\end{equation}
(whence
\be\label{def-posi-2}
\mathbf{a}_{(1)}= \mathbf{x}_{a}+ \mathbf{x}_{1a}c^{-2})
\ee
and
\begin{equation}\label{def-velo}
 \mathbf{v}_{a}:= \dot{\mathbf{a}}_{(1)}  = \dot{\mathbf{x}}_{a}+
\dot{\mathbf{x}}_{1a}c^{-2}.
\end{equation}
The general form of the equations of motion of (any of) the mass centers is got by integrating the local equations of motion inside the domain $\mathrm{D}_a$ occupied by the corresponding body $(a)$. At the order zero, this gives the Newtonian equation of motion:
\begin{equation}\label{masscent-ord0}
  M_a\,\ddot{a}^i= \int_{\mathrm{D}_a} \rho U ^{(a)}_{,i}\, \dd V.
\end{equation}
Here we use Fock's \cite{Fock1964} decomposition of $U$ into an external part and a self part
\begin{eqnarray}\label{FockDecompos}
U & =  & U  ^{(a)} + u _{a} \\
U  ^{(a)}( {\bf x} ) & := & G \sum_{b \neq a} \int_{\mathrm{D}_b} \, \rho({\bf y}) \dd V({\bf y})/\abs{{\bf x-y} } \\
u _{a}( {\bf x} ) & := & G \int_{\mathrm{D}_a} \, \rho({\bf y}) \dd V({\bf y})/\abs{{\bf x-y}}.
\end{eqnarray}

For the order one, the integration inside body $(a)$ of Eq. (\ref{i-ord1}) gives \cite{A35}
\begin{equation}\label{masscent-ord1}
  M_a^1\,\ddot{a}_1^i+\dot{I}^{ai}=J^{ai}+K^{ai}
\end{equation}
with
\begin{equation}\label{Iai}
  I^{ai}:= \int_{\mathrm{D}_a} [p+\rho(\mathbf{u}^2/2+\Pi+3U)]u^i \dd {\mathsf V}
\end{equation}
\begin{equation}\label{Jai}
  J^{ai}:= \int_{\mathrm{D}_a} (\sigma_1 U_{,i}+\rho V_{1,i}) \dd {\mathsf V}
\end{equation}
and
\begin{eqnarray}\nonumber
  K^{ai}& := & \int_{\mathrm{D}_a}[-2Up_{,i}+pU_{,i}+ \rho u^i\partial _TU + \rho {\bf u}^2 U_{,i}]\dd {\mathsf V} \\
& = & 3 \int_{\mathrm{D}_a}pU_{,i}+ \int_{\mathrm{D}_a} \rho u^i\partial _TU + \int_{\mathrm{D}_a} \rho {\bf u}^2 U_{,i}\dd {\mathsf V} \nonumber \\ \label{Kai}
& := &  \quad \ 3K_1^{ai}\quad  + \quad   K_2^{ai} \qquad + \qquad  \ K_3^{ai}.
\end{eqnarray}
Using Eq.~(\ref{delta-a}), Eq.~(\ref{masscent-ord1}) allows us to compute the 1PN correction to the acceleration of the 1PN mass centers ${\bf a}_{(1)}$:
\be\label{delta-addot}
\ddot{{\bf a}}_{(1)} - \ddot{{\bf a}} = \ddot{\mathbf{x}}_{1a}c^{-2}= \frac{-\dot{{\bf I}}^{a}+{\bf J}^{a}+{\bf K}^{a}-M_a^1 \ddot{{\bf a}}}{c^2 M_a}+ O(c^{-4}).
\ee
The integrals ${\bf I}^{a}:=(I^{ai})_{i=1,2,3}$, ${\bf J}^{a}$, $M^1_a$, and ${\bf K}^{a}$, entering this equation, are calculated in Appendix \ref{Integrals-eta}: Eqs. (\ref{Idotai_SS}), (\ref{Ja-La-M1addota}), (\ref{Lai-external-explicit}), (\ref{M1a-explicit}), (\ref{K1}), (\ref{K2}), and (\ref{K3}). Reporting those results into Eq. (\ref{delta-addot}), we get first after elementary simplifications:

\begin{eqnarray} \label{EMMC-ETGv2-1}
\ddot{\mathbf{x}}_{1a} & = & \dot{{\bf a}} G \sum_{b\neq a} 2M_b \frac{{\bf n}_{ab}}{r_{ab}^2} \left( 2 \dot{{\bf a}}-\dot{{\bf b}}\right ) \\
& & + G \sum_{b\neq a}  \frac{\left  [-\left (\alpha _b+\alpha _a\frac{M_b}{M_a}\right )+ M_b\left (\frac{3}{2}({\bf n}_{ab}.\dot{{\bf b}})^2-\frac{\dot{{\bf b}}^2}{2}\right ) \right ]{\bf n}_{ab} }{r_{ab}^2} \nonumber \\
& & -G\sum _{b\neq a} M_b \frac{({\bf n}_{ab}.\dot{{\bf b}})\dot{{\bf b}}}{r_{ab}^2} + G \sum_{b\neq a}\frac{M_b}{2} \frac{ \left ({\bf n}_{ab}.\ddot{{\bf b}}\right ){\bf n}_{ab} -\ddot{{\bf b}}}{r_{ab}} \nonumber \\
& & + G\sum_{b\neq a} M_b  \frac{\mathbf{x}_{1b}\mathbf{-}\mathbf{x}_{1a}+
  3\left[(\mathbf{x}_{1a}\mathbf{-}\mathbf{x}_{1b})\mathbf{.}\mathbf{n}_{ab}
  \right]\mathbf{n}_{ab}} {r_{ab}^3} \nonumber \\
& & + \left [\frac{\dot{{\bf a}}^2}{2}-3 U^{(a)}({\bf a}) -\left (\frac{\dot{{\bf a}}^2}{2}+3 U^{(a)}({\bf a}) \right)_{T=0} -\frac{5T_a +\frac{41}{3}\varepsilon _a}{M_a}\right ]\ddot{{\bf a}} \nonumber \\
& & + 3 \frac{\xi _a}{M_a} \boldsymbol{\omega}_{a}\wedge \dot{{\bf a}}+O\left(c^{-2}\right) +o(\eta^3). \nonumber
\end{eqnarray}
In this equation, we may use the expression (\ref{spherical-phi-ext}) of the external potential $U^{(a)}$, which corresponds with the expression
\be\label{spherical-phi-ext-grad}
\ddot{{\bf a}} = \nabla U^{(a)}({\bf a})+ O(\eta ^4) = G \sum_{b \neq a} \left(\frac{ -{\bf n}_{ab}}{r_{ab}^2}\right )  M_b +O(\eta ^4)
\ee
for the Newtonian acceleration (\ref{masscent-ord0}). Therefore, using also (\ref{alpha-M1a}) and (\ref{M1a-explicit}), we may group together line 2 and line 5 in Eq. (\ref{EMMC-ETGv2-1}) as
\be\label{line2+line5}
G \sum_{b \neq a} \left(\frac{ -{\bf n}_{ab}}{r_{ab}^2}\right ) \kappa _{a b}
\ee
with
\bea
\kappa _{a b} & = & M_b\left [\dot{{\bf a}}^2 + \dot{{\bf b}}^2 -\frac{3}{2}({\bf n}_{ab}.\dot{{\bf b}})^2 -4U^{(a)}({\bf a})-U^{(b)}({\bf b})\right ] \nonumber \\
& & + M_b\left [\dot{{\bf b}}_{T=0}^2+3 U^{(b)}({\bf b})_{T=0} + \frac{58 \varepsilon _a -4T_a}{3M_a}+ \frac{17 \varepsilon _b +11T_b}{3M_b}\right ] \cdot
\eea
We may also use the equivalent of (\ref{spherical-phi-ext-grad}) for the Newtonian acceleration $\ddot{{\bf b}}$ of any of the other bodies, which intervenes in line 3 of Eq. (\ref{EMMC-ETGv2-1}). Moreover, we found previously \cite{A36} that, if we define the vector radius in terms of the full 1PN positions (\ref{def-posi-2}):
\be\label{def-Nab-Rab}
R_{ab} := \abs{ {\bf a}_{(1)} - {\bf b}_{(1)} }, \qquad \mathbf{N}_{ab} := \frac{ {\bf a}_{(1)} - {\bf b}_{(1)}}{R_{ab}}
\ee
instead of defining it in terms of the zero-order positions, as in (\ref{def-n0ab-r0ab}), then we have the following 1PN expansion:
\be\label{relation-Nab-nab}
\frac{\mathbf{N}_{ab} } {R_{ab}^2} = \frac{\mathbf{n}_{ab} }{r_{ab}^2} +\frac{1}{c^2} \left\{\frac{ \mathbf{x}_{1a} -\mathbf{x}_{1b} -3\left[(\mathbf{x}_{1a}\mathbf{-}\mathbf{x}_{1b})\mathbf{.}\mathbf{n}_{ab}
  \right]\mathbf{n}_{ab}}{r_{ab}^3}  \right\} +O\left(c^{-4}\right).
\ee
We recognize in the second term the coefficient of $-GM_b$ in line 4 of Eq. (\ref{EMMC-ETGv2-1}). Until now (including also the Appendix), we used the ``spherical'' expressions (\ref{spherical-phi-ext}) and (\ref{spherical-phi-ext-grad}) of the Newtonian potential only to compute the 1PN correction to the acceleration. If we use it also for the Newtonian main term, we can thus group Eq. (\ref{spherical-phi-ext-grad}) and line 4 of Eq. (\ref{EMMC-ETGv2-1}) into the total acceleration $\ddot{{\bf a}}_{(1)}$ as
\be\label{Newt-accel-Nab-Rab}
G \sum_{b \neq a} \left(\frac{ -{\bf N}_{ab}}{R_{ab}^2}\right )  M_b .
\ee
For the other terms in the 1PN correction (\ref{EMMC-ETGv2-1}) to the acceleration, we can use indifferently $r_{ab}$ or $R_{ab}$, and $\dot{{\bf a}}= \dot{{\bf x}}_a$ or $\dot{{\bf a}}_{(1)}= {\bf v}_a$, since their difference is $O(c^{-2})$, hence makes an $O(c^{-4})$ difference, which we neglect, in the total acceleration $\ddot{{\bf a}}_{(1)}$.\\

Using all of these remarks, we deduce from (\ref{EMMC-ETGv2-1}) that
\begin{eqnarray} \label{EMMC-ETGv2-2}
\ddot{{\bf a}}_{(1)} & = & \sum_{b\neq a}  -\frac{GM_b {\bf N}_{ab}}{R_{ab}^2}\left \{ 1+\delta_b + \frac{1}{c^2} \left [{\bf v}_a^2 +{\bf v}_b^2-\frac{3}{2}({\bf N}_{ab}.{\bf v}_b )^2 - 4\sum_{d\neq a} \frac{GM_d}{R_{ad}} \right.\right. \nonumber \\
& & \left.\left. -\sum_{d \neq b} \frac{GM_d}{R_{bd}} \left ( 1+ \frac{R_{ab}}{2 R_{bd}}\mathbf{N}_{ab}.\mathbf{N}_{bd}\right ) \right ] \right \} + \sum_{b\neq a} \frac{GM_b}{2R_{ab}} \sum_{d \neq b} \frac{GM_d}{c^2 R^2_{bd}}\mathbf{N}_{bd} \nonumber \\
& & +\sum_{b\neq a} \frac{GM_b}{c^2 R^2_{ab}} \left[\left(4\mathbf{N}_{ab}.{\bf v}_a - 2\mathbf{N}_{ab}.{\bf v}_b \right) {\bf v}_a - (\mathbf{N}_{ab}.{\bf v}_b){\bf v}_b \right ]\nonumber \\
& & + 3 \frac{\xi _a}{c^2 M_a} \boldsymbol{\omega}_{a}\wedge \dot{{\bf a}} + O\left(c^{-4}\right) +o(\eta^3)
\end{eqnarray}
with
\be
\delta _b := \frac{1}{c^2} \left [({\bf v}_b^2)_{T=0}+3 U^{(b)}({\bf b})_{T=0} + \frac{58 \varepsilon _a -4T_a}{3M_a}+ \frac{17 \varepsilon _b +11T_b}{3M_b}\right ] \cdot
\ee
Obviously, $\delta _b$ disappears from Eq. (\ref{EMMC-ETGv2-2}) if we redefine the Newtonian masses as $M'_a := M_a(1+\delta _a)$. Equation (\ref{EMMC-ETGv2-2}) can be readily compared with the usual writing of the Lorentz-Droste(-Einstein-Infeld-Hoffmann) equations \cite{PoissonWill2014, LDEIH_Wiki}. It can equivalently be written in the somewhat simpler form
\begin{eqnarray} \label{EMMC-ETGv2-3}
\ddot{{\bf a}}_{(1)} & = & \sum_{b\neq a}  -\frac{GM_b {\bf N}_{ab}}{R_{ab}^2}\left \{ 1+\delta_b + \frac{1}{c^2} \left [{\bf v}_a^2 +{\bf v}_b^2-\frac{3}{2}({\bf N}_{ab}.{\bf v}_b )^2 -4U^{(a)}({\bf a}) \right.\right. \nonumber \\
& & \left.\left. -U^{(b)}({\bf b}) \right ] \right \} + \sum_{b\neq a} \frac{GM_b}{c^2}  \frac {\left( \mathbf{N}_{ab}.\dot{{\bf v}}_b \right )\mathbf{N}_{ab} - \dot{{\bf v}}_b}{2R_{ab}} \nonumber \\
& & +\sum_{b\neq a} \frac{GM_b}{c^2 } \frac{\left(4\mathbf{N}_{ab}.{\bf v}_a - 2\mathbf{N}_{ab}.{\bf v}_b \right) {\bf v}_a - (\mathbf{N}_{ab}.{\bf v}_b){\bf v}_b}{R^2_{ab}} \nonumber \\
& & + 3 \frac{\xi _a}{c^2 M_a} \boldsymbol{\omega}_{a}\wedge \dot{{\bf a}} + O\left(c^{-4}\right) +o(\eta^3).
\end{eqnarray}

\section{Conclusion}\label{Conclusion}

In this paper, the equations of motion of the mass centers of a weakly-gravitating system of $N$ bodies have been derived according to the asymptotic scheme of PN approximation \cite{A23}, for v2 of the investigated scalar theory: Eq. (\ref{EMMC-ETGv2-2}) or equivalently Eq. (\ref{EMMC-ETGv2-3}). As in the case of GR \cite{A36}, the asymptotic scheme makes it clear that the internal structure of the bodies does influence their motion, here through the appearance in this equation of the structure parameters $\varepsilon _a$, $T_a$ and $\xi_a$ ($a=1,...,N$). In an operational view, only the $\xi_a$'s do effectively play a role, in the sense that the other parameters $\varepsilon _a$ and $T_a$ enter only through the combination $\delta_a$, which can be eliminated or rather ``hidden'' thanks to a redefinition of the Newtonian masses $M_a$. Even then, and even if one leaves apart the term involving $\xi_a$, this equation of motion differs from the Lorentz-Droste(-Einstein-Infeld-Hoffmann) equations. As to the spin term with $\xi_a$, this is a ``self'' acceleration that is present even if there is only one body, as with the different spin term that enters the equation derived according to the asymptotic scheme for GR \cite{A36}. It is likely that using either of these equations obtained with the asymptotic scheme (the present equations as also those obtained for GR \cite{A36}) should lead to a non-negligible difference as compared with using the standard, Lorentz-Droste(-Einstein-Infeld-Hoffmann) equations. However, for the relatively short interval of time for which we have precise observations, it might be the case that an adjustment of parameters would lead to a reasonable fit of the data --- especially, if one would consider ``direct'' data, meaning ones for which the celestial mechanics based on the standard equations has not been used to ``reduce'' (to correct) the rough data.

\appendix
\section{Appendix: The Integrals ${\bf I}^{a}$, ${\bf J}^{a}$ and ${\bf K}^{a}$ for Well-Separated, Rigidly-Rotating Bodies} \label{Integrals-eta}

The integrals ${\bf I}^{a}$ and ${\bf J}^{a}$ are modifications of integrals denoted in the same way, which have been computed for v1 \cite{A25-26,A32} and for GR in the harmonic gauge \cite{A36}.
\subsection{Integral ${\bf I}^{a}$ and its Time Derivative}

By comparing Eq. (\ref{Iai}) with Eq. (4.8) in Ref. \cite{A25-26}, one sees that
\be\label{Iai_new_vs_old}
I^{ai} _{\mathrm{v2}} = I^{ai} _{\mathrm{v1}} + 2 \int_{D_a} \rho U u^i\,\dd V
\ee
with (Eqs. (A10) and (A11) in Ref. \cite{A25-26}, see also Ref. \cite{A36}, Eqs. (A4) and (A5))
\be \label{I1ai}
I^{ai}_{\mathrm{v1}} = I^{ai}_2 + (M_a \dot{{\bf a}}^2/2 + 2 T_a + 4 \varepsilon_a) \dot{a}^i + (\dot{a}^k \Omega^{(a)}_{lk}I^{(a)}_{jl} + 2T_{aj}+ 4 \varepsilon_{aj})\Omega^{(a)}_{ji} +O(\eta ^{7/2})
\ee
\be\label{I2ai}
I^{ai}_2 := \int_{\mathrm{D}_a} \rho U^{(a)} u^i \, \dd V = M_a \dot{a}^i U^{(a)}({\bf a}) + I^{(a)}_{jk} \Omega^{(a)}_{ki} U^{(a)}_{,j}({\bf a}) + O(\eta ^{7/2})
\ee
where \cite{Fock1964}
\begin{equation}\label{def-epsa-epsai}
  \varepsilon_a:= \int_{\mathrm{D}_a} \rho u_a \dd V/2, \quad \varepsilon_{aj} := \int_{\mathrm{D}_a} \rho u_a (x^j-a^j) \dd V/2
\end{equation}
\begin{equation}\label{def-Iaij-Omegaa}
I^{(a)}_{ij}:= \int_{\mathrm{D}_a} \rho(x^i-a^i)(x^j-a^j) \dd V, \quad \Omega_a := \Omega^{(a)}_{ik} \Omega^{(a)}_{jk}(x^i-a^i)(x^j-a^j)/2
\ee
\begin{equation}\label{def-Ta-Taj}
T_a:= \int_{\mathrm{D}_a} \rho \Omega_a \dd V,\quad T_{aj} :=  \int_{\mathrm{D}_a} \rho \Omega_a (x^j-a^j) \dd V.
\end{equation}
We have
\be\label{int rho U ui}
\int_{D_a} \rho U u^i\,\dd V = \int_{D_a} \rho U^{(a)} u^i\,\dd V + \int_{D_a} \rho u_a u^i\,\dd V
\ee
with, from (\ref{u-rigid}) and (\ref{def-epsa-epsai}),
\be\label{int rho ua ui}
\int_{D_a} \rho u_a u^i\,\dd V = 2 (\varepsilon_a \dot{a}^i + \varepsilon_{aj}\Omega^{(a)}_{ji}).
\ee
From (\ref{Iai_new_vs_old}), (\ref{I1ai}), (\ref{I2ai}), (\ref{int rho U ui}), and (\ref{int rho ua ui}), we get:
\bea\label{Iai_v2}
I^{ai} _{\mathrm{v2}} & = & (M_a \dot{{\bf a}}^2/2 + 2 T_a + 8 \varepsilon_a ) \dot{a}^i + 3M_a \dot{a}^i U^{(a)}({\bf a}) + (2T_{aj}+ 8 \varepsilon_{aj})\Omega^{(a)}_{ji} \nonumber \\
& & + 3  I^{(a)}_{jk} \Omega^{(a)}_{ki} U^{(a)}_{,j}({\bf a}) + \dot{a}^k \Omega^{(a)}_{lk} I^{(a)}_{jl} \Omega^{(a)}_{ji} +O(\eta ^{7/2}).
\eea

To compute the {\it 1PN correction} to the final equations of motion (but not necessarily to compute the main, Newtonian part), we assume that the bodies have spherical symmetry. See Ref. \cite{A36} for a discussion of this assumption. It implies that
\be\label{epsaj_Taj_Iajk_SS}
\varepsilon_{aj} = T_{aj} =0, \quad I^{(a)}_{jk} = \gamma_a \delta_{jk}
\ee
with
\be \label{gamma_a}
 \gamma_a := \frac{4\pi}{3}\int_0^{r_a} \, r^4 \rho_a(r) \dd r
\ee
(here $\rho_a(r):= \rho({\bf x}),\ r:=\abs{{\bf x}-{\bf a}},\ {\bf x} \in \mathrm{D}_a$). This allows us to rewrite (\ref{Iai_v2}) as
\bea\label{Iai_SS}
I^{ai} & = & (M_a \dot{{\bf a}}^2/2 + 2 T_a + 8 \varepsilon_a ) \dot{a}^i + 3M_a \dot{a}^i U^{(a)}({\bf a}) \nonumber \\
& & + 3  \gamma_a \Omega^{(a)}_{ji} U^{(a)}_{,j}({\bf a}) + \dot{a}^k \gamma_a \Omega^{(a)}_{jk} \Omega^{(a)}_{ji} +O(\eta ^{7/2}).
\eea
We differentiate this with respect to the time $T$. For well-separated bodies such that the Newtonian velocity field is that of a rigid rotation, and with a Newtonian density $\rho$ that is spherical (or only quasi-spherical in the sense of Ref. \cite{A36}), the rate of the (Newtonian) angular rotation velocity is $O(\eta^3)$ (\cite{A36}, Eq. (B8)). It follows from this and Eqs. (\ref{adot-eta}) and (\ref{omega-eta}) that, in Eq. (\ref{Iai_SS}), the last term before the remainder gives an $o(\eta^3)$ contribution to $\dot{I}^{ai}$, which we neglect. It also follows from that ``Newtonian sphericity'' assumption that the external Newtonian potential is given by
\bea \label{spherical-phi-ext}
U^{(a)}({\bf x}) & = & \sum_{b \neq a} \frac{G M_b } {\abs{{\bf x}-{\bf b}}} \quad ({\bf x} \in \mathrm{D}_a) \\
 \frac{\dd}{\dd T}[U^{(a)}({\bf a})] & = & - \sum_{b \neq a} G M_b  \frac{(b^j-a^j)(\dot{b}^j-\dot{a}^j)}{{\abs{{\bf a}-{\bf b}}}^3} \\
 \frac{\dd}{\dd T}[U^{(a)}_{,j}({\bf a})] & = & O(\eta ^{7/2}).
\eea
Still, the rigid motion of the bodies implies that the integral $\varepsilon_a$ [Eq. (\ref{def-epsa-epsai})] is time-independent, and also (\cite{A36}, Eq. (A13)), that, for well-separated bodies, $\dd T_a/\dd T = O(\eta ^{7/2})$. Accounting for the foregoing remarks, we obtain from (\ref{Iai_SS}):
\bea\label{Idotai_SS}
\dot{I}^{ai} & = & \left [ M_a \left (\frac{\dot{{\bf a}}^2}{2} + 3U^{(a)}({\bf a}) \right) + 2 T_a + 8 \varepsilon_a \right ] \ddot{a}^i + M_a ( \dot{{\bf a}}.\ddot{{\bf a}})\dot{a}^i  \nonumber \\ & & -3 M_a \dot{a}^i \sum_{b \neq a} G M_b  \frac{(b^j-a^j)(\dot{b}^j-\dot{a}^j)}{{\abs{{\bf a}-{\bf b}}}^3} + o(\eta^3).
\eea

\subsection{Integral ${\bf J}^{a}$}

This integral [Eq. (\ref{Jai})] has been calculated in Ref. \cite{A25-26} for v1, and the remainders in that calculation have been evaluated with the asymptotic framework for good separation in Ref. \cite{A32}. In Ref. \cite{A36}, it has been calculated for GR in the harmonic gauge. The difference between the two calculations, and with the present one, regards only the definition of the 1PN correction to the active mass density, $\sigma _1$ (involving the 1PN correction $\rho_1$ to the rest-mass density). In the case considered: spherically symmetric Newtonian density (and rigid rotation), we have (Ref. \cite{A36}, Eqs. (A35), (A44)-(A46)):
\begin{eqnarray} \label{Ja-La-M1addota}
\frac{ {\bf J}^{a}-{\bf L}^{a}-M_a^1 \ddot{{\bf a}} } {M_a} & = & G\sum_{b\neq a} \left[\alpha_b +(\alpha_a-M^1_a)\frac{M_b}{M_a} \right]\left(\frac{-{\bf n}_{ab}}{(r_{ab})^2}\right) \\
&  & + G\sum_{b\neq a} \frac{M_b}{(r_{ab})^3}\left[\mathbf{x}_{1b}\mathbf{-}\mathbf{x}_{1a}+
  3\left((\mathbf{x}_{1a}\mathbf{-}\mathbf{x}_{1b})\mathbf{.}\mathbf{n}_{ab}
  \right)\mathbf{n}_{ab}\right] +O(\eta^4) \nonumber
\end{eqnarray}
where
\footnote{\
To obtain Eq. (\ref{Ja-La-M1addota}), the following expression of $\beta_{aj} := \int_{\mathrm{D}_a} \sigma_1({\bf x}) (x^j - a^j) \, \dd V ({\bf x})$ is used:
\be\label{def-beta-aj}
\beta_{aj} = M^1_a (a_1^j-a^j) + \eta_{aj} +O(\eta)
\ee
(Eqs. (A27) and (A34) in Ref. \cite{A36}), with moreover $\eta_{aj}=0$ if the Newtonian density $\rho$ is spherical inside body $(a)$ (see after Eq. (A37) in Ref. \cite{A36}). It is easy to check, adapting the line of calculation in  Ref. \cite{A36}, that (\ref{def-beta-aj}) is valid also for the investigated theory, despite the different formula for $\sigma _1$, here Eq. (\ref{sigma_1}).
}
\be\label{def-alpha-a}
\alpha_a := \int_{\mathrm{D}_a} \sigma _1 \, \dd V
\ee
\be\label{def-n0ab-r0ab}
r_{ab} := \abs{{\bf a} - {\bf b}} := \abs{ {\bf x}_{a} - {\bf x}_{b} }, \qquad \mathbf{n}_{ab} := \frac{{\bf a} - {\bf b}}{r_{ab}}
\ee
and
\bea\label{Lai-external-explicit}
{\bf L}^{a} & = & \frac{GM_a}{2}\sum_{b\neq a} \ M_b \left\{ \frac{\left(\mathbf{n}_{ab}\mathbf{.}\ddot{\mathbf{b}}\right)\mathbf{n}_{ab}
  -\ddot{\mathbf{b}}}{r_{ab}}
  +\frac{\left[3(\mathbf{n}_{ab}\mathbf{.}\dot{\mathbf{b}})^2-
  \dot{\mathbf{b}}^2\right]\mathbf{n}_{ab}
  -2\left(\mathbf{n}_{ab}\mathbf{.}\dot{\mathbf{b}}\right)\dot{\mathbf{b}} } {(r_{ab})^2}\right \} \nonumber \\
&& -\frac{2}{3}\varepsilon_a \ddot{{\bf a}} + O(\eta^4).
\eea
Using the fact that the formula (\ref{sigma_1}) for $\sigma _1$ differs from v1 only by the presence of $-\rho U$ in v2 instead of $+\rho U$ in v1, we deduce easily the value of $\alpha _a$ for v2 from that for v1, Eq. (A8) in Ref. \cite{A32}: for v2,
\be\label{alpha-M1a}
\alpha_a = M^1_a +M_a \left[ \dot{{\bf a}}^2 -U^{(a)}({\bf a}) \right] + \frac{8}{3}T_a -\frac{1}{3}\varepsilon _a +O(\eta ^3).
\ee
Also, noting that (\ref{rho_1}) is identical to the corresponding equation for GR in the harmonic gauge, Eq. (2.50)$_3$ in Ref. \cite{A36}, the calculation of $M^1_a$ there is valid for v2, thus (Eq. (A32) in Ref. \cite{A36})
\bea\label{M1a-explicit}
M^1_a & = & \left ( M^1_a \right )_{T=0} = \left (\int_{\mathrm{D}_a} \rho\left (\frac{{\bf u}^2}{2} + 3U \right) \right )_{T=0} \nonumber \\
& = & M_a \left[\frac{\dot{{\bf a}}^2}{2}+3U^{(a)}({\bf a}) \right]_{T=0} + T_a+6\varepsilon _a +O(\eta ^3).
\eea

\subsection{Integral ${\bf K}^a$}
This integral is the sum of three terms, Eq. (\ref{Kai}). The first integral, $K^{ai}_1$, has been calculated in Ref. \cite{A32}, Eq. (A25) (though it was denoted $K^{ai}_2$ there):
\be\label{K1}
K^{ai}_1 := \int_{\mathrm{D}_a} pU_{,i} = \frac{1}{3}(\varepsilon_a - 2T_a)U^{(a)}_{,i}({\bf a}).
\ee

To compute $K^{ai}_2$, we use the decomposition (\ref{FockDecompos}). Assuming that the Newtonian density $\rho$ is spherical inside body $(a)$, we have
\be
u_a({\bf x},T) = \int_{\abs{{\bf x}-{\bf a}(T)}} ^\infty \frac{G \mu_a(r)}{r^2}\,\dd r,\qquad \mu_a(r):= 4\pi \int_0 ^r \rho_a(s) s^2 \dd s
\ee
\bea
\partial_T u_a({\bf x},T) & = & \partial_T \int_{\abs{{\bf x}-{\bf a}(T)}} ^\infty \frac{G \mu_a(r)}{r^2} \,\dd r \nonumber \\
& = & -\left ( \frac{G \mu_a(r)}{r^2}\right )_{r=\abs{{\bf x}-{\bf a}(T)}}\partial_T \,\abs{{\bf x}-{\bf a}(T)} \nonumber \\
& = & -\frac{G \mu_a(r)}{r^2} \left( -\dot{a}^k \frac{x^k-a^k}{\abs{{\bf x}-{\bf a}}} \right ) = \frac{G \mu_a(r)}{r^2} \dot{a}^k n^k
\eea
(denoting henceforth $n^j := (x^j-a^j)/\abs{{\bf x}-{\bf a}}$ inside $\mathrm{D}_a$), whence, using (\ref{u-rigid}) and noting $\dd \omega$ the element of solid angle:
\bea\label{int rho d_T ua ui}
\int _{\mathrm{D}_a} \rho \partial_T u_a u^i \dd V & = & \int \dd \omega \int_0 ^{r_a} r^2  \dd r \rho_a(r) \frac{G \mu_a(r)}{r^2} \dot{a}^k n^k (\dot{a}^i+\Omega_{ji} rn^j)  \nonumber \\
& = & \frac{4 \pi}{3} \Omega_{ji} \dot{a}^j \int_0 ^{r_a} r \rho_a(r) G \mu_a(r) \dd r \nonumber \\
& = & \xi_a \left ( \boldsymbol{\omega}_{a}\wedge \dot{{\bf a}} \right )^i
\eea
with
\be
\xi_a = -\frac{4 \pi}{3} \int_0 ^{r_a} \rho_a(r) \frac{\dd u_a}{\dd r} r^3 \dd r.
\ee
(We used the well-known integrals $\int n^k \dd \omega =0,\ \int n^j n^k \dd \omega =\frac{4 \pi}{3} \delta_{j k}$.) On the other hand, from (\ref{spherical-phi-ext}), we obtain
\bea \label{d_T_spherical-phi-ext}
\partial_T U^{(a)}({\bf x}) & = & \sum_{b \neq a} G M_b \frac{\dot{b}^k (x^k-b^k)} {\abs{{\bf x}-{\bf b}}^3}
\eea
whence, using (\ref{u-rigid}):
\bea \label{int rho ui d_T_spherical-phi-ext}
\int _{\mathrm{D}_a} \rho \partial_T U^{(a)} u^i \dd V & = & \sum_{b \neq a} G M_b \dot{a}^i \dot{b}^k \int_{\mathrm{D}_a} \rho \frac{x^k-b^k} {\abs{{\bf x}-{\bf b}}^3} \dd V \nonumber \\
& & + \sum_{b \neq a} G M_b \int_{\mathrm{D}_a} \rho \Omega^{(a)}_{ji} rn^j \dot{b}^k\frac{x^k-b^k} {\abs{{\bf x}-{\bf b}}^3} \dd V.
\eea
We can write using the Taylor formula:
\be
\frac{x^k-b^k} {\abs{{\bf x}-{\bf b}}^3} = \frac{a^k-b^k} {\abs{{\bf a}-{\bf b}}^3} + O(\eta^3).
\ee
From this, using (\ref{adot-eta}) and (\ref{omega-eta}), one finds that the second integral in (\ref{int rho ui d_T_spherical-phi-ext}) is $O(\eta^4)$, and that
\bea \label{int rho ui d_T_spherical-phi-ext-2}
\int _{\mathrm{D}_a} \rho \partial_T U^{(a)} u^i \dd V & = & \sum_{b \neq a} G M_b \dot{a}^i \dot{b}^k \int_{\mathrm{D}_a} \rho \frac{a^k-b^k} {\abs{{\bf a}-{\bf b}}^3} \dd V +O(\eta^4) \nonumber \\
& = &  \dot{a}^i  M_a \sum_{b \neq a} G M_b  \frac{({\bf a}-{\bf b}).\dot{{\bf b}}}{\abs{{\bf a}-{\bf b}}^3}+O(\eta^4).
\eea
Thus, by summing (\ref{int rho d_T ua ui}) and (\ref{int rho ui d_T_spherical-phi-ext-2}):
\be\label{K2}
{\bf K}^{a}_2 = \xi_a \boldsymbol{\omega}_{a}\wedge \dot{{\bf a}} + \dot{{\bf a}} M_a \sum_{b \neq a} G M_b  \frac{({\bf a}-{\bf b}).\dot{{\bf b}}}{\abs{{\bf a}-{\bf b}}^3}+O(\eta^4).
\ee

To compute $K^{ai}_3$ [Eq. (\ref{Kai})], we write using (\ref{u-rigid}) and (\ref{FockDecompos}):
\be\label{K3_1}
K^{ai}_3 = \int_{\mathrm{D}_a} \rho \left [\dot{{\bf a}}^2 + 2 \dot{a}^k \Omega^{(a)}_{jk}(x^j-a^j)+2\Omega _a \right ] \left (U^{(a)}_{,i} +u_{a,i} \right ) \dd {\mathsf V}.
\ee
With the assumed spherical symmetry, we have
\be\label{u_a,i}
u_{a,i} = \frac{\dd u_a}{\dd r}\frac{x^i-a^i} {\abs{{\bf x}-{\bf a}}} = \frac{\dd u_a}{\dd r} n^i.
\ee
Therefore,
\be
\int_{\mathrm{D}_a} \rho \left (\dot{{\bf a}}^2 +2\Omega _a \right ) u_{a,i}  \dd {\mathsf V} = \left (\dot{{\bf a}}^2 +2\Omega _a \right ) \int _0 ^{r_a}  \rho_a(r) \frac{\dd u_a}{\dd r} r^2 \dd r \int n^i \dd \omega =0
\ee
so that
\bea
& & \int_{\mathrm{D}_a} \rho \left [\dot{{\bf a}}^2 + 2 \dot{a}^k \Omega^{(a)}_{jk}(x^j-a^j)+2\Omega _a \right ] u_{a,i} \dd {\mathsf V} \nonumber \\
&& = 2 \dot{a}^k \Omega^{(a)}_{jk} \int _0 ^{r_a} \rho_a(r) r \frac{\dd u_a}{\dd r} r^2 \dd r \int n^i n^j \dd \omega \nonumber \\
& & = 2  \dot{a}^k \Omega^{(a)}_{jk} \int _0 ^{r_a} \rho_a(r) r^3 \frac{\dd u_a}{\dd r} \dd r \times \frac{4 \pi}{3} \delta_{i j} = 2 \xi_a \left ( \boldsymbol{\omega}_{a}\wedge \dot{{\bf a}}\right )^i. \label{K3_2}
\eea
On the other hand, since $U^{(a)}_{,i}=O(\eta^2)$, we obtain from (\ref{omega-eta})
\bea\label{K3_3}
\int_{\mathrm{D}_a} \rho \left [\dot{{\bf a}}^2 + 2 \dot{a}^k \Omega^{(a)}_{jk}(x^j-a^j)+2\Omega _a \right ] U^{(a)}_{,i} \dd {\mathsf V} & = & \int_{\mathrm{D}_a} \rho \dot{{\bf a}}^2  U^{(a)}_{,i} \dd {\mathsf V} + o(\eta^3) \nonumber \\
& = & M_a \dot{{\bf a}}^2  U^{(a)}_{,i}({\bf a}) + o(\eta^3) .
\eea
In view of (\ref{K3_1}), we get by summing (\ref{K3_2}) and (\ref{K3_3})
\be\label{K3}
{\bf K}^{a}_3 = 2 \xi_a \boldsymbol{\omega}_{a}\wedge \dot{{\bf a}} + M_a \dot{{\bf a}}^2  \nabla U^{(a)}({\bf a}) + o(\eta^3) .
\ee

\end{document}